\documentstyle[12pt]{article}
\makeatletter
\newcommand{\singlespacing}{\let\CS=\@currsize\renewcommand{\baselinestretch}
{1.0}\tiny\CS}
\newcommand{\doublespacing}{\let\CS=\@currsize\renewcommand{\baselinestretch}
{1.5}\tiny\CS}

\begin{document}

\begin{center}
{\large {\bf OPTIMAL CLONING AND NO SIGNALING}} 

\vspace{0.4cm}
{\bf Sibasish Ghosh}\footnote{res9603@isical.ac.in},~ {\bf Guruprasad Kar},~ {\bf Anirban Roy}\footnote{res9708@isical.ac.in} 

\vspace{0.2cm}
{\it Physics and Applied Mathematics Unit}\\
{\it Indian Statistical Institute}\\
{\it 203 B. T. Road, Calcutta--700 035}\\
{\it India}.

\vspace{1.3cm}
ABSTRACT

It is shown that no signaling
constraint generates the whole class of 1 $\rightarrow$ 2 optimal quantum
cloning machines of single qubits
\end{center}

\vspace{2.5cm}
\flushleft{PACS NO. 03.65.Bz}

\newpage
 
Perfect cloning ({\it i.e.}, copying) of an unknown quantum mechanical state is
known to be impossible, as shown by Wooters and Zurek. This is valid for pure
quantum mechanical states \cite{wooters}, and well as for mixed states \cite{barnum}.
Bu\v{z}ek and Hillery have provided a universal 1 $\rightarrow$ 2 cloning machine,
which produces two identical but imperfect copies of an arbitrary single qubit
state \cite{buzek1}. Bru{\ss} et. al. \cite{bruss} have shown that this symmetric
(as the two copies produced are identical) universal cloning machine of Bu\v{z}ek 
and Hillery is optimal. 
%Recently Cerf 
%\cite{cerf1, cerf2} has provided a family of
%quantum cloning machines each of which produces two imperfect copies of the
%input qubit state, using a Pauli channel for the overall input-to-output
%operation for each copy. 
Recently Cerf \cite{cerf2} has provided a concept of asymmetric quantum cloning
when the two output states of the cloner are not identical, but at the same time,
these two output states are specifically related to the input. 
The cloning operation presented in \cite{cerf2} is universal for qubits,
{\it i.e.}, the fidelity of cloning does not depend on the input qubit state. 
Bu\v{z}ek et. al. \cite{buzek2} have provided a
universal 1 $\rightarrow$ 2 cloning network for assymmetric cloning using
local unitary operations and controlled NOT (C-NOT) operations, where the
input state is a single qubit, and the optimal symmetric cloning
machine of Bu\v{z}ek and Hillery \cite{buzek1} is reproduced. 

In a very interesting way, Gisin \cite{gisin}
has connected the 1 $\rightarrow$ 2 symmetric cloning operation of qubits with no
signaling property (which states that superluminal signaling is impossible in
quantum mechanics). In this letter, we shall reproduce the result of Bu\v{z}ek et.
al. \cite{buzek2} using no signaling condition. And our derivation shows that
the univerasal 1 $\rightarrow$ 2 assymmetric cloning machine of Bu\v{z}ek et. al.
\cite{buzek2} is optimal (described bellow).   

$a_0$ corresponds to the original single qubit, $a_1$ corresponds to the
blanck copy (which is also in a single qubit state), and $b_1$ corresponds to
the machine of the cloning process. Let ${\rho}_{a_0}^{\rm in}~ (\vec{m}) =
(1/2) (I + \vec{m} . \vec{\sigma})$ be the density matrix of the input single qubit state (which is
unknown, as the Bloch vector $\vec{m} = ({\rm sin} \theta {\rm cos} \phi,
{\rm sin} \theta {\rm sin} \phi, {\rm cos} \theta)$ is unknown) entering into 
the asymmetric quantum cloning machine (AQCM). We want to clone
(asymmetrically) this qubit universally, {\it i.e.}, input-state independently
({\it i.e.}, independent of the Bloch vector $\vec{m}$), in such a way that the
density matrices of 
the two clones ${\rho}_{a_j}^{\rm out}~ (\vec{m})$ ($j = 0, 1$) at the output
of the AQCM are of the forms\\
\begin{equation}
\label{outputs}
{\rho}_{a_j}^{\rm out}~ (\vec{m}) = s_j {\rho}_{a_j}^{\rm in}~ (\vec{m}) +
\frac{1 - s_j}{2} I,
\end{equation}
(for $j = 0, 1$) where $I$ is the $2 \times 2$ identity matrix. Equation
(\ref{outputs}) is referred as the isotropy condition. Obviously here $0 \le s_0,
s_1 \le 1$. For symmetric QCM, $s_0 =
s_1$. Let ${\rho}_{a_0 a_1}^{\rm out}~ (\vec{m})$ be the two qubit output density matrix 
of the AQCM, obtained after employing the trace operation on the machine states
in the output pure state $| \psi \rangle _{a_0 a_1 {\rm machine}}^{\rm out}~ (\vec{m})$,
obtained by applying the asymmetric cloning operation on ${\rho}_{a_0}^{\rm in}~ (\vec{m})$. 
In full generality, ${\rho}_{a_0 a_1}^{\rm out}~ (\vec{m})$ can be written as\\
\begin{equation}
\label{2qout}
{\rho}_{a_0 a_1}^{\rm out}~ (\vec{m}) = \frac{1}{4} \left(I \times I + s_0 \vec{m} .
\vec{\sigma} \otimes I + s_1 I \otimes \vec{m} . \vec{\sigma} + \sum_{j, k = x,
y, z}^{}~ t_{jk} {\sigma}_j \otimes {\sigma}_k \right).
\end{equation} 
The AQCM will be universal if it acts similarly on all input states, {\it
i.e.}, if 
\begin{equation}
\label{universal}
{\rho}_{a_0 a_1}^{\rm out}~ ({\bf R} \vec{m}) = U({\bf R}) \otimes U({\bf R})
{\rho}_{a_0 a_1}^{\rm out}~ (\vec{m}) {U({\bf R})}^{\dag} \otimes {U({\bf R})}^{\dag},
\end{equation}
where ${\bf R} \equiv {\bf R}(\vec{n}, \alpha)$ represents an arbitrary
rotation (in $SO(3)$) about an axix along the unit vector $\vec{n}$ through an
angle $\alpha$ of the
Bloch vector $\vec{m}$, and $U({\bf R}) \equiv e^{-i \frac{\alpha}{2} \vec{n}
. \vec{\sigma}}$ is the corresponding $2 \times 2$
unitary operation (it is in $SU(2)$) acting on the two 2-dimensional Hilbert
spaces corresponding to the two qubits $a_0$ and $a_1$. As a consequence of
this property (given by equation (\ref{universal})), we see that (see
\cite{gisin}) ${\rho}_{a_0 a_1}^{\rm out}~ (\vec{m})$ is invariant under
rotation of $\vec{m}$, {\it i.e.},\\
\begin{equation}
\label{rotation}
\left[e^{i \alpha \vec{m} . \vec{\sigma}} \otimes e^{i \alpha \vec{m} .
\vec{\sigma}}, {\rho}_{a_0 a_1}^{\rm out}~ (\vec{m}) \right] = 0~ {\rm for}~
{\rm all}~ {\rm real}~ \alpha.
\end{equation}
Equation (\ref{rotation}) imposes the following conditions on the parameters
$t_{jk}$ :\\
%\end{document}
\begin{equation}
\label{tjk}
\left.
\begin{array}{lrr}
- m_z t_{xy} + m_y t_{xz} - m_z t_{yx} + m_y t_{zx} &=& 0,\\
m_z t_{xx} - m_x t_{xz} - m_z t_{yy} + m_y t_{zy} &=& 0,\\
- m_y t_{xx} + m_x t_{xy} - m_z t_{yz} + m_y t_{zz} &=& 0,\\   
m_z t_{xx} - m_z t_{yy} + m_y t_{yz} - m_x t_{zx} &=& 0,\\  
m_z t_{xy} + m_z t_{yx} - m_x t_{yz} - m_x t_{zy} &=& 0,\\ 
m_z t_{xz} - m_y t_{yx} + m_x t_{yy} - m_x t_{zz} &=& 0,\\ 
- m_y t_{xx} + m_x t_{yx} - m_z t_{zy} + m_y t_{zz} &=& 0,\\   
- m_y t_{xy} + m_x t_{yy} + m_z t_{zx} - m_x t_{zz} &=& 0,\\   
- m_y t_{xz} + m_x t_{yz} - m_y t_{zx} + m_x t_{zy} &=& 0.
\end{array}
\right\}
\end{equation}
%\end{document}
Partcularly, for $\vec{m} = (0, 0, 1) \equiv  \uparrow$, we have
$t_{xx}^{\uparrow} = t_{yy}^{\uparrow}$, $t_{xy}^{\uparrow} =
-t_{yx}^{\uparrow}$ and $t_{yz}^{\uparrow} = t_{zy}^{\uparrow} =
t_{zx}^{\uparrow} = t_{xz}^{\uparrow} = 0$.\\
For $\vec{m} = (0, 0, -1) \equiv \downarrow$, we have $t_{xx}^{\downarrow} = t_{yy}^{\downarrow}$, $t_{xy}^{\downarrow} =
-t_{yx}^{\downarrow}$ and $t_{yz}^{\downarrow} = t_{zy}^{\downarrow} =
t_{zx}^{\downarrow} = t_{xz}^{\downarrow} = 0$.\\
For $\vec{m} = (1,0, 0) \equiv  \rightarrow$, we have $t_{yy}^{\rightarrow} = t_{zz}^{\rightarrow}$, $t_{yz}^{\rightarrow} =
-t_{zy}^{\rightarrow}$ and $t_{zx}^{\rightarrow} = t_{xz}^{\rightarrow} =
t_{xy}^{\rightarrow} = t_{yx}^{\rightarrow} = 0$.\\  
And for $\vec{m} = (-1,0, 0) \equiv  \leftarrow$, we have $t_{yy}^{\leftarrow} = t_{zz}^{\leftarrow}$, $t_{yz}^{\leftarrow} =
-t_{zy}^{\leftarrow}$ and $t_{zx}^{\leftarrow} = t_{xz}^{\leftarrow} =
t_{xy}^{\leftarrow} = t_{yx}^{\leftarrow} = 0$.\\  
Our motivation is to find out bounds on $s_0$ and $s_1$ (or some sort of relation between
them), using the conditions imposed on $t_{jk}$'s by equation (\ref{tjk}), the
no signaling condition (to be described bellow), and the positive
semi-definiteness of the density matrices ${\rho}_{a_0 a_1}^{\rm out}~
(\vec{m})$ for each Bloch vector $\vec{m}$. 

If it would have been possible to distinguish between different mixtures that can be
prepared at a distance ({\it e.g.}, between ${\rho}_{a_0 a_1}^{\rm out}~ (\vec{m}) +
{\rho}_{a_0 a_1}^{\rm out}~ (- \vec{m})$ and ${\rho}_{a_0 a_1}^{\rm out}~ (\vec{m^{\prime}}) +
{\rho}_{a_0 a_1}^{\rm out}~ (- \vec{m^{\prime}})$), then non-locality in
quantum mechanics could be used for signaling ({\it i.e.}, superluminal
signaling) through that distance, and hence we would reach at a contradiction
between quantum mechanics and relativity \cite{shimony}. Thus we have to
maintain no signality, which imposes that the mixtures ${\rho}_{a_0 a_1}^{\rm out}~ (\vec{m}) +
{\rho}_{a_0 a_1}^{\rm out}~ (- \vec{m})$ and ${\rho}_{a_0 a_1}^{\rm out}~ (\vec{m^{\prime}}) +
{\rho}_{a_0 a_1}^{\rm out}~ (- \vec{m^{\prime}})$ (of the output states),
corresponding to the indistinguishable mixtures $(1/2) (I + \vec{m} .
\vec{\sigma}) + (1/2) (I - \vec{m} . \vec{\sigma})$ and $(1/2) (I + \vec{m^{\prime}} .
\vec{\sigma}) + (1/2) (I - \vec{m^{\prime}} . \vec{\sigma})$ respectively (of
the input states), are themselves indistinguishable \cite{gisin}. So, without
loss of generality, we can write\\
\begin{equation} 
\label{nosignaling}
{\rho}_{a_0 a_1}^{\rm out}~ (0, 0, 1) +
{\rho}_{a_0 a_1}^{\rm out}~ (0, 0, -1) = {\rho}_{a_0 a_1}^{\rm out}~ (1, 0, 0) +
{\rho}_{a_0 a_1}^{\rm out}~ (-1, 0, 0).
\end{equation}
Using equations (\ref{tjk}) and (\ref{nosignaling}), we get the following
expression for ${\rho}_{a_0 a_1}^{\rm out}~ (\uparrow)$, where $\uparrow = (0,
0, 1)$ :
$${\rho}_{a_0 a_1}^{\rm out}~ (\uparrow) = \frac{1}{4} [I \otimes I + s_0
{\sigma}_z \otimes I + s_1 I \otimes {\sigma}_z + t \left({\sigma}_x \otimes
{\sigma}_x + {\sigma}_y \otimes {\sigma}_y + {\sigma}_z \otimes {\sigma}_z
\right)$$
\begin{equation}
\label{uparrow}
+  t_{xy} \left({\sigma}_x \otimes {\sigma}_y - {\sigma}_y \otimes
{\sigma}_x \right)],
\end{equation}
where $t = t_{xx}^{\uparrow} = t_{yy}^{\uparrow} = t_{zz}^{\uparrow}$ and
$t_{xy} = t_{xy}^{\uparrow}$ are both real quantities. The (real) eigen values of ${\rho}_{a_0 a_1}^{\rm
out}~ (\uparrow)$ are given by\\
\begin{equation}
\label{eigenvalues}
\frac{1}{4} \{1 + t \pm (s_0 + s_1)\},~~ \frac{1}{4} [1 - t \pm \{4 t^2 + 4
t_{xy}^2 + (s_0
- s_1)^2\}^{1/2}].
\end{equation}
All these eigen values must be non-negative, and so we must have\\
\begin{equation}
\label{inequality1}
s_0 + s_1 \le 1 + t,
\end{equation}
\begin{equation}
\label{inequality2}
(s_0 - s_1)^2 + 4 t_{xy}^2 \le (1 + t)(1 - 3t),
\end{equation}
\begin{equation}
\label{inequality3}
-1 \le t \le \frac{1}{3}.
\end{equation} 
From equation (\ref{inequality1}) we see that maximum values of both $s_0$ and
$s_1$ will occur when\\
\begin{equation}
\label{maximum}
s_0 + s_1 = 1 + t.
\end{equation}
So, using equation (\ref{maximum}), we get from equation (\ref{inequality2})
that\\
\begin{equation}
\label{relation}
{s_0}^2 + {s_1}^2 + s_0 s_1 - s_0 - s_1 + t_{xy}^2 \le 0.
\end{equation}
The optimal symmetric cloning machine of Bu\v{z}ek and Hillery \cite{buzek1}
(where $s_0 = s_1 = 2/3$) will be reproduced here if we take $t_{xy} = 0$, and
then condition (\ref{relation}) is exactly equation (11) of \cite{buzek2}. And from equation
(\ref{maximum}) we see that the relation (\ref{relation}) has to be satisfied
by the reduction factors $s_0, s_1$ of an optimal AQCM, which implies that the
AQCM of Bu\v{z}ek et. al. \cite{buzek2} is optimal.    

No signaling constraint was used \cite{gisin} to derive the optimality of the
(universal) symmetric cloning machine of Bu\v{z}ek and Hillery \cite{buzek1}, and in this paper the same constraint
has been used to find out the optimality of the (universal) asymmetric cloning
machine of Bu\v{z}ek et. al. \cite{buzek2}. 

%{\noindent {\bf Acknoledgement}:} 

\end{document}